\documentclass[preprint,aps,showpacs]{revtex4}
\usepackage{graphicx}

\usepackage{amsmath}
\usepackage{amsxtra}
\usepackage{amstext}
\usepackage{amssymb}
\usepackage{latexsym}
\usepackage{dsfont}

\newcommand{\be}{\begin{equation}}
\newcommand{\ee}{\end{equation}}
\newcommand{\bn}{\begin{eqnarray}}
\newcommand{\en}{\end{eqnarray}}

\begin{document}

\title{How can the Standard model Higgs and also the extensions of the Higgs to Yukawa's scalars  
 be interpreted in the {\it spin-charge-family} theory and to what predictions about the 
 Higgs does this theory lead?}

\author{Norma Susana Manko\v c  Bor\v stnik}
\affiliation{Department of Physics, FMF, University of Ljubljana,
Jadranska 19, Ljubljana, 1000}

%\maketitle

\begin{abstract}
This contribution is to show how does the  {\em spin-charge-family} theory~\cite{NF,norma,pikanorma} 
interpret the assumptions of the {\em standard model}, and those  extensions   
of this model, which are trying to see the Yukawa couplings as scalar fields with the family
(flavour) charges in the fundamental representations of the group. 
The purpose of these contribution is i.) to try to understand why the {\em standard model} works so well, 
although its assumptions look quite artificial, and ii.) how do predictions of the 
{\em spin-charge-family} theory about the measurements of the scalar fields differ from predictions 
of the {\em standard model}, which has only one scalar field - the Higgs - and also from its more or less 
direct extensions with Yukawas as the scalar dynamical fields with the family charge in the fundamental
or anti-fundamental representation of group. 
% This last has to be comment in the paper.
\end{abstract}


\maketitle

\section{Introduction}
\label{Introduction}

When the {\em standard model} of the elementary particles and fields was 
proposed more than  35 years ago it offered an elegant new step in understanding 
the origin of fermion and boson fields and the interactions among them. 

It postulated: 
{\bf i.)}   The existence  of the family members - coloured quarks and colourless 
leptons, both left and right handed, the left handed members distinguishing from the right 
handed ones in  the weak and hyper charges. 
{\bf ii.)}  The existence of the gauge fields to the observed charges of the family members. 
{\bf iii.)} The existence of the scalar field, the Higgs, which takes care of the masses of 
weak gauge fields and fermions and is chosen to have the charges, the weak one and the hyper one, 
in the fundamental representations of the weak group, just like fermions, to "dress right handed" 
family members with the weak and the right hyper charge and to interact with the gauge fields 
to bring masses only to $Z_{m}$ and $W^{\pm}_{m}$. 
{\bf iv.)}  The existence of the families.
{\bf v.)}   The existence of the Yukawa couplings, taking care of the masses of fermions, 
 together with the Higgs. 
 
The {\em standard model}  never has the ambition to explain its own assumptions, 
leaving the explanation of the open questions to the next step of the theory. 

While all the so far observed {\it fermions} are {\it spinors} with the {\it charges in the fundamental 
representations of the charge groups} and all the {\it bosons} are {\it vectors} in the 
{\it adjoint representations with respect 
to all the charges}, are the  {\it Higgs} field and its anti-Higgs {\it scalars} with the {\it charges in 
the fundamental representations of the weak charge group}. Therefore, quite a strange object, which reminds us 
on a supersymmetric particle (but it is not because it does not fit the so called R parity requirements).

Although the {\em standard model} leaves many questions unanswered, yet it is, without any doubt, a very 
efficient effective theory: There is so far no experiment which would help to show the next step 
beyond the {\em standard model}. 

In the literature there are several proposals trying to  go beyond the standard model, most of them 
just extending the ideas of the {\em standard model}, like:
i.)  A tiny extension is the inclusion of the right handed neutrinos into the family. 
ii.) The $SU(3)$ group is assumed to describe -- not explain -- the existence of three families. 
iii.) Like Higgs has the charges in the fundamental representations of the groups, also
Yukawas are assumed to be scalar fields, in the bi-fundamental %(fundamental for left handed family members and 
%anti-fundamental for the right handed ones) 
representation of the $SU(3)$ group.
vi.) Supersymmetric theories assuming the existence of partners to the existing  fermions 
and bosons, with charges in the opposite representations, adjoint for fermions and fundamental for bosons.

Let me  comment on the {\em standard model} as it is seen  from the point of view of the 
{\em  spin-charge-family} theory~\cite{NF,norma,pikanorma}, which is, to my knowledge, the only proposal in 
the literature which is offering the mechanism for generating families, not just postulating families 
with an appropriate choice of a family group. 

I shall comment also on the extensions of the Higgs scalar fields to the Yukawa dynamical fields carrying 
the family triplet or anti-triplet charges, depending on the handedness of fermions and  
distinguishing also among the family members. These extensions~\cite{Georgi,giudice,belen} 
are meant as an effective stage to help to learn from the experiment something about the nature of 
the scalar fields. 

Although  effective interactions can have quite unexpected shapes and yet can be very useful  
(as it is the case, for example, with the by experiments suggested spin-spin interaction in several 
models in the solid state and the liquid crystal physics where the effective interaction of the 
electromagnetic origin among many electrons and nuclei can be effectively expressed with the spin-spin 
interaction) yet it is hard to accept that effective theories of the type where the $SU(3)$ 
groups are used to describe  the family quantum number, introducing the  scalar dynamical fields 
with the family charges in the fermionic like representations, 
can make useful predictions for new experiments, where searches depend strongly on the proposed 
theories behind. To my understanding at this stage of physics a new 
more general understanding of fermion and boson fields is needed.

Let me in what follows briefly present the {\it standard model} and its extension with the family group 
added and the {\it spin-charge-family} theory and comment with the last one the {\it standard model} 
and its extensions.

\section{The most urgent open questions in the elementary particle physics}
\label{questions}

To my understanding any new step in theoretical explanation  of the {\it standard model} 
assumptions must answer the following most urgent open questions: % without these answers  
%there can hardly be any progress in the  elementary particle physics and cosmology, are: 
\begin{itemize}
\item  What is the origin of families? 
  How many families there are at all?
\item What is the origin of the scalar  fields (the Higgs)? 
 Where do their  masses   (the Higgs mass) and correspondingly  
the masses of the gauge fields originate? 
\item What is the  origin of the fermions masses,  
 where do Yukawa couplings originate?
\item Where does the dark matter originate?
\end{itemize}
 There are also several other questions which may not be so urgently   answered, like:
Where do dark energy originate?  
What is the origin of charges, and correspondingly of the gauge fields?  And several others.

The {\em spin-charge-family} theory~\cite{norma,pikanorma,Portoroz03,gmdn,GN} 
seems  promising in answering these and several other open questions. 

\section{The assumptions of the standard model% and of its extensions with the family gauge group
}
\label{smextensions}

The standard model assumes  before the electroweak break: 

\noindent
i.) Three massless families of fermions, each one containing quarks and leptons in the fundamental  
(spinor) representations with respect to the spin and the charges as presented in table~\ref{Table I.}.   
\begin{table}
%{\tiny%
\begin{center}
\begin{tabular}{|r c r r r r |}
\hline
$\alpha$& hand- &  weak  & hyper  & colour & elm  \\
        & edness& charge & charge & charge & charge\\

name        &$ -4i S^{03} S^{12}$&$ \tau^{13}$  &$ Y$          &               &$Q$\\
\hline\hline
$u^{i}_{L}$  &$               -1 $&$ \frac{1}{2}$&$ \frac{1}{6}$& colour triplet&$\frac{2}{3}$\\
\hline
$d^{i}_{L}$  &$  -1              $&$-\frac{1}{2}$&$ \frac{1}{6}$& colour triplet&$-\frac{1}{3}$\\
\hline
$\nu^{i}_{L}$&$  -1              $&$ \frac{1}{2}$&$-\frac{1}{2}$&  colourless   &$0$  \\
\hline
$e^{i}_{L}  $&$  -1              $&$-\frac{1}{2}$&$-\frac{1}{2}$&  colourless   &$-1$ \\
\hline\hline
$u^{i}_{R}  $&$   1              $& weakless     &$ \frac{2}{3}$& colour triplet&$\frac{ 2}{3}$\\
\hline
$d^{i}_{R}  $&$   1              $& weakless     &$-\frac{1}{3}$& colour triplet&$-\frac{1}{3}$\\
\hline
$\nu^{i}_{R}$&$   1              $& weakless     &$ 0          $&  colourless   &$0$           \\
\hline
$e^{i}_{R}  $&$   1              $& weakless     &$ -1         $&  colourless   &$-1$ \\
\hline\hline
\end{tabular}
  \end{center}
\caption{\label{Table I.} Properties of one family of quarks and leptons, right handed $\nu$ is included.
$\tau^{13}$ defines the third component of the weak charge, $Y$ the hyper charge,  $Q= Y + \tau^{13}$ is the 
electromagnetic charge. 
} %}
\end{table}
%\end{itemize}
Members of each of  the three ($i=1,2,3$) before the electroweak break massless (so far observed) 
families are assumed to fit the experimental data.
Each family contains the left handed weak charged quarks and the right 
handed weak chargeless quarks,   belonging  
to the colour triplet $(1/2,1/(2\sqrt{3}))$, $(-1/2,1/(2\sqrt{3}))$, $(0,-1/(\sqrt{3})) $.    
%}
%
The existence of the corresponding anti-fermion  is assumed for each family member of any family. 
(Let me add that in the {\it spin-charge-family} theory  all these properties of one family follow 
from the simple starting action, from where also family members follow.) 
%\item

\noindent
ii.) Three massless gauge fields of the hyper charge ($Y$), the weak charge ($\vec{\tau}^1$) and the 
colour charge $(\vec{\tau}^3)$ are assumed as presented in table~\ref{Table II.}. 
\begin{table}
%{\tiny%
\begin{center}
\begin{tabular}{|r|c|r|r|r|c|}
\hline
name & hand-         &   weak  & hyper  &  colour & elm \\
     & edness        &   charge& charge &  charge & charge\\  
\hline\hline
  hyper photon       &$ 0$              &$ 0$     &  $0$      & colourless  &$0$\\
\hline
  weak bosons        &$ 0$              & triplet &  $0$      & colourless  & triplet \\
\hline
  gluons             &$ 0$              &$0$      &  $0$      & colour octet&$0$ \\
\hline\hline
\end{tabular}
  \end{center}
\caption{\label{Table II.} Vector  gauge fields in $d=(1+3)$, 
the gauge fields of the hyper, weak and colour charges.
} %}
\end{table}
They all are vectors in $d=(1+3)$,  in the adjoint 
representations with respect to the weak, colour and hyper charges, $Q= \tau^{13} + Y$. 
(In the {\it spin-charge-family} theory  all these gauge fields follow  
from the simple starting action with  vielbeins and two kinds of the spin connection fields as the 
only gauge fields follow.)

\noindent
iii.) The Higgs field,  the scalar in $d=(1+3)$, is assumed to be in the fundamental representation 
with respect to the  charges. %$P_{R}= (-1)^{2 s + 3 B + L} =1.$
\begin{table}
%{\tiny%
\begin{center}
\begin{tabular}{|r|c|r|r|r|c|}
\hline
name       & hand-         &   weak        &  hyper         &  colour   &  elm \\
           & edness        &   charge      & charge         &  charge   &  charge\\  
\hline\hline
%name &handedness         & weak charge&hypercharge & colour charge & elm charge\\
%\hline
Higgs$_{u}$&$ 0$           &$  \frac{1}{2}$&$ \frac{1}{2}  $& colourless&$  1$\\
\hline
Higgs$_{d}$&$ 0$           &$- \frac{1}{2}$&$ \frac{1}{2}  $& colourless&$  0$\\                                        
\hline\hline
\end{tabular}
  \end{center}
\caption{\label{Table III.} Higgs is a scalar field in $d=(1+3)$, with the charges 
in the fundamental representation 
 of the charge groups. %Before the electroweak break is massless. 
 } %}
\end{table} 
And there is the anti-Higgs. 
(In the {\it spin-charge-family} theory  there are several scalar fields, all with the 
charges in the adjoint representations with respect to all the charge groups, which
 follow from the simple starting action.)

\section{The assumptions of the spin-charge-family theory}
\label{mine}

There exist two kinds of the Clifford algebra objects.  
{\bf i.)} The Dirac one, $\gamma^a$, is used to describe the spin of fermions. 
In the Kaluza-Klein-like theories the spin in higher dimensions together with the angular moments
manifest, after the appropriate breaks of symmetries, as the charges  and the spin in (1+3).
{\bf ii.)} The second kind, I call it $\tilde{\gamma}^a$, anti-commuting with the Dirac one, describes 
families, since it forms the equivalent representations with respect to the first one (what means that 
 $S^{ab}$ apply on each family in the same way). 

Accordingly there exist beside vielbeins two kinds of spin-connection fields in gravity.
The {\it spin-charge-family} theory is assuming the simplest action in $ d= (1 + 13)$: 
\begin{eqnarray}
S            \,  &=& \int \; d^dx \; E\;{\mathcal L}_{f} +  
 \int \; d^dx \; E\; (\alpha \,R + \tilde{\alpha} \, \tilde{R}),
               \end{eqnarray}
with the Lagrange densities for spinors and for gauge fields as follows 
\begin{eqnarray}
{\mathcal L}_f &=& \frac{1}{2}\, (\bar{\psi} \, \gamma^a p_{0a} \psi) + h.c., \quad 
p_{0a }         = f^{\alpha}{}_a p_{0\alpha} + \frac{1}{2E}\, \{ p_{\alpha}, E f^{\alpha}{}_a\}_-, 
\nonumber\\  
   p_{0\alpha} &=&  p_{\alpha}  - 
                    \frac{1}{2}  S^{ab} \omega_{ab \alpha} - 
                    \frac{1}{2}  \tilde{S}^{ab}   \tilde{\omega}_{ab \alpha},                   
\nonumber\\ 
R              &=&  \frac{1}{2} \, \{ f^{\alpha [ a} f^{\beta b ]} \;(\omega_{a b \alpha, \beta} 
- \omega_{c a \alpha}\,\omega^{c}{}_{b \beta}) \} + h.c. \;, 
\nonumber\\
\tilde{R}      &=& \frac{1}{2}\,   f^{\alpha [ a} f^{\beta b ]} \;(\tilde{\omega}_{a b \alpha,\beta} - 
\tilde{\omega}_{c a \alpha} \tilde{\omega}^{c}{}_{b \beta}) + h.c.\;.  
\label{wholeaction}
\end{eqnarray}
The action~(\ref{wholeaction}) manifests in $d=(1+3)$, after the breaks of symmetries   (chosen to lead 
to the measured properties of fermions and bosons) and before the electroweak break,  
the  families of left handed weak charged quarks and leptons and right handed weakless quarks and 
leptons. Just as it is assumed by the {\it standard model}, except that an additional (broken) $U(1)$ charge
exists and that there are also right handed neutrinos. The right handed neutrinos have nonzero value 
of this additional $U(1)$ charge. And there are four (rather then assumed three) families. 

Before the electroweak break there are  therefore four massless families of quarks and leptons 
and four massive families. After the electroweak break there are two groups of four 
massive families. The off diagonal and diagonal mass matrices of family members are 
determined on the tree level by the dynamical scalar fields, the gauge fields of the 
charges originating in $\tilde{\gamma}^a$. To the diagonal terms also the scalar fields 
originating in $\gamma^s$, $s \ge 5$ contribute. In loop corrections both fields start to 
contribute coherently. The scalar dynamical fields determine also masses of $Z_{m}$  and 
$W^{\pm}_{m}$. For more information 
the reader is kindly asked to see the refs.~\cite{NF,AN} in this proceedings and in the 
references therein.

All scalar and vector bosons have all the charges, except the weak charge in the adjoint representations 
of the charge gauge groups. All the scalar fields contributing to the masses of the lower four families 
are doublets with respect to the weak charge. 
All fermions have all the charges in the fundamental representations with respect to all 
the groups. The family quantum numbers are here included. 

The only assumptions in this theory are that before the breaks of the symmetries the 
simplest action for fermions and bosons is taken, in which the only internal degrees of freedom 
are the two kinds of a spin, determined by $\gamma^a$ and $\tilde{\gamma}^{a}$, respectively, 
requiring accordingly besides the vielbeins the two kinds of the spin connection fields,
 and that the breaks (phase transitions) lead to measured phenomena.

After the breaks the first kind of spin manifests in $(1+3)$ the spin and all the charges,
while the second one manifests the families.

\section{Predictions of the spin-charge-family theory}
\label{predictions}

The {\it spin-charge-family} theory predictions are  so far:

i.) There are two groups of four families in the low energy regime. The fourth of the lowest 
four families waits to be measured. The fifth stable family constitute the dark matter.
More about these topics can be found in the refs.~\cite{NF,AN,GN}.

ii.) The scalar fields, which are responsible for the mass matrices of fermions and correspondingly 
for their masses and mixing matrices, manifest effectively as the Higgs field and the 
Yukawa couplings. Effectively means that they are several dynamical fields with the 
family charges in the adjoint representations of while with respect to the weak charge they 
are doublets, like it is the {\it standard model} Higgs.

Accordingly, I expect that future searching for scalar fields will show up that  there are several 
dynamical scalar fields, not just one, and that each one couples differently to different family members. 
More about this topic can be read in the ref.~\cite{NF}.

iii.) What is evident from the {\it spin-charge-family} theory is that besides the known gauge 
fields also the scalar fields are the interaction fields. Accordingly, since they effectively 
are representing also Yukawas, indeed also Yukawas are the interacting fields, although in the 
{\it standard model} hidden within the Higgs.

iv.) There are no supersymmetry in the low energy regime. Even Higgs, which is very close to 
a supersymmetric particle being a scalar with the charges in the fundamental representations 
of the charge groups, is only effectively working as it does. In the {\it spin-charge-family-theory}
it is the operator $\gamma^0 \gamma^s, \, s=7,8,\,$ which transforms the right  handed family members 
into the corresponding left handed ones, "doing the dressing job" of the  Higgs and of $\gamma^0$.

iv.) There are also several other predictions, not yet enough studied to be commented here.

\section{How are the standard model and its extensions seen from the point of view of the 
spin-charge-family theory?}
\label{smexstension}

As already discussed above, the {\it standard model} assumed the  existence of a scalar field 
which has charges in the fundamental representations of the charged groups. As I have shown in the 
ref.~\cite{NF}, one can define an operator which does effectively the Higgs job.  

The extensions of the {\it standard model}~\cite{Georgi,giudice,belen}, where also Yukawas are 
taken as dynamical fields under the assumption that there are several family  $SU(3)$ groups, one 
for the left handed quarks and two for the right handed quarks and similarly (not quite because of 
the right handed neutrinos) for the leptons. Then there are scalar dynamical fields assumed, again 
(under the influence of the Higgs) in the fundamental representations of these family groups. 
%In order that the mass term is a scalar the "left handed" Yukawas are in the 
%triplet representation and couple to the left handed family members, while the  "right handed" 
%Yukawas, which couple to the right handed family members, are in the anti-triplet representations 
%of the family group. 
%I can hardly see that these "strange objects", a kind of a supersymmetric fields again,
%could help at all to predict the right experiments, which would have a chance to measure 
%the {\it spin-charge-family} theory scalars, which have all the charges in the adjoint 
%representation of the charge groups. 

\section{Conclusions}

Let me conclude this discussion by again pointing out that the {\it spin-charge-family} theory 
does answer  the questions, which are, to my understanding, the most urgent to be answered for 
any new successful step beyond the {\it standard model}. I doubt that trying to explain only one 
of  the "urgent open questions" presented above, can bring a new insight into the assumptions of the 
{\it standard model}. 

The {\it spin-charge-family} theory offers not only the answer to the question why  we have 
more than one family, and how many there are, it explains also the origin of the Higgs and the 
Yukawa couplings, of the charges and the gauge fields. 

Correspondingly it predicts the fourth and the stable fifth families. There are also other open 
questions, which the theory is offering the answers for. 

According to these predictions there is no supersymmetric particles at the low energy regime.
But, without doubts, there are additional families and several additional dynamical fields - interactions.
Work is in progress.

\end{document}